\documentclass[preprint,preprintnumbers,amsmath,amssymb]{revtex4}
\usepackage{graphicx}
\usepackage{epstopdf}
\usepackage{dcolumn}
\usepackage{bm}
\usepackage{color}
\usepackage{array,multirow,graphicx}

\begin{document}

\title{Dirac-Electron Behavior for Spin-Up Electrons in Strongly Interacting Graphene on Ferromagnetic Mn$_5$Ge$_3$}

\author{Elena Voloshina$^{1,2,}$\footnote{Corresponding author. E-mail: voloshina@shu.edu.cn} and Yuriy Dedkov$^{1,}$\footnote{Corresponding author. E-mail: dedkov@shu.edu.cn}}

\affiliation{\mbox{$^1$Department of Physics, Shanghai University, 99 Shangda Road, 200444 Shanghai, China}}
\affiliation{\mbox{$^2$Physical and Theoretical Chemistry, Freie Universit\"at Berlin, 14195 Berlin, Germany}}

\date{\today}

\begin{abstract}

An elegant approach on the synthesis of graphene on the strong ferromagnetic (FM) material Mn$_5$Ge$_3$ is proposed via intercalation of Mn in the graphene-Ge(111) interface. According to the DFT calculations, graphene in this strongly interacting system demonstrates the large exchange splitting of the graphene-derived $\pi$ band. In this case only spin-up electrons in graphene preserve the Dirac-electron-like character in the vicinity of the Fermi level and the K point, whereas such behavior is not detected for the spin-down electrons. This unique feature of the studied gr/FM-Mn$_5$Ge$_3$ interface which can be prepared on the semiconducting Ge can lead to its application in spintronics.\\

This document is the unedited author's version of a Submitted Work that was subsequently accepted for publication in J. Phys. Chem. Lett., copyright $\copyright$ American Chemical Society after peer review. To access the final edited and published work see doi: 10.1021/acs.jpclett.9b00893.

\end{abstract}

\maketitle

The recent progress in the intensive studies of the graphene-metal (gr-M) interfaces placed these systems in the forefront of the 2D-materials and particularly of the graphene-related research~\cite{Geim:2009,Batzill:2012,Dedkov:2015kp}. This situation is due to not only the perspectives of the mass-production of the high-quality graphene on metals~\cite{Bae:2010,Ryu:2014fo}, but mainly because of the intriguing properties of the graphene-metal interface itself and that different gr-M systems are used as a playground in fundamental studies of the properties of 2D systems and as a basis for the preparation of more complex 2D-based heterosystems~\cite{Meng:2012ee,Xie:2018bv,Ehlen:2019jj}. Among the gr-M systems the distinct attention is paid to the graphene-ferromagnet (gr-FM) interfaces and significant understanding of the properties of these systems has been achieved~\cite{Dedkov:2010jh,Voloshina:2011wa,Usachov:2015kr}. However, there are some factors which limit the further progress in the fundamental studies and in applications of these interesting objects. For example, the strong interaction at the gr-Ni(111) or gr-Co(0001) interfaces leads to the high degree of the  $\pi-3d$ hybridization of the valence band states at the interface between graphene and FM metal and, as a consequence, graphene loses its free-standing character and the carriers in the vicinity of the Fermi ($E_F$) level cannot be described as mass-less Dirac particles~\cite{Karpan:2007,Dedkov:2010jh,Voloshina:2011wa,Usachov:2015kr}. Yet, the same reason causes appearance of induced magnetic moments in graphene~\cite{Weser:2010,Matsumoto:2013eu}. In this case the states at $E_F$, which are significant for the description of the charge and spin transport properties of the junction, mainly have FM $3d$ character, substantially changing the band dispersion, states character and Fermi velocity of the carriers. Following works demonstrate the possibility to decouple graphene from the underlying FM material via intercalation of different species~\cite{Batzill:2012,Dedkov:2015kp}, giving a route to restore the free-standing character of graphene, however losing the possibility of the induced spin polarization in a graphene layer~\cite{Voloshina:2011NJP,Vita:2014aa,Omiciuolo:2014dn}.

Further progress in the graphene synthesis on different substrates and desire to implement graphene in the modern semiconductor technology led to the recent discovery of the graphene synthesis on the catalytically active semiconducting Ge substrates of different orientations~\cite{Lee:2014dv}. This and subsequent studies demonstrate that Ge$(110)$ and Ge$(111)$ planes are more suitable for the graphene growth~\cite{Kiraly:2015kaa,Wang:2016ca,Tesch:2018hm}, whereas in case of Ge(001) a significant surface faceting under graphene layer~\cite{Pasternak:2016ec,Dabrowski:2017gr} was found limiting further technological processing of the later interface. The recent progress in this area is mainly focused on the studies of the growth mechanism of graphene on Ge surfaces and the studies of the electronic structure of gr-Ge interface or its modification are very rare~\cite{Dabrowski:2017gr,Tesch:2018hm,Zhou:2018ji,Ahn:2018fj}.

In the present work, we propose the elegant approach to prepare new gr-FM interface, namely gr-Mn$_5$Ge$_3$, which combines several well approved steps used in the graphene-related studies. We performed density functional theory (DFT) calculations of the lattice-mismatched gr-Mn$_5$Ge$_3$(0001) interface in the super-cell geometry and it is computationally shown that in this strongly interacting system the graphene $\pi$ states undergo strong exchange split leaving only spin-up electrons in the vicinity of $E_F$ which preserve the Dirac-electron-like character for the carriers with the linear energy band dispersion. At the same time, for the spin-down electrons a series of localized interface states is formed that strongly reduces the carriers' mobility for this spin channel. Such difference for the electron mobilities \textbf{can} lead to the effective spin-filtering effect in such single-layer graphene-based junction. We believe that the described simple approach, which can be used for the preparation of this gr-FM interface, will be extremely useful in the future studies and applications, where 2D graphene as well as FM and semiconductor materials are included in one heterosystem. Details of DFT calculations and additional data are presented as the Supporting Information.

\begin{figure}[t]
\centering
\includegraphics[width=0.57\columnwidth]{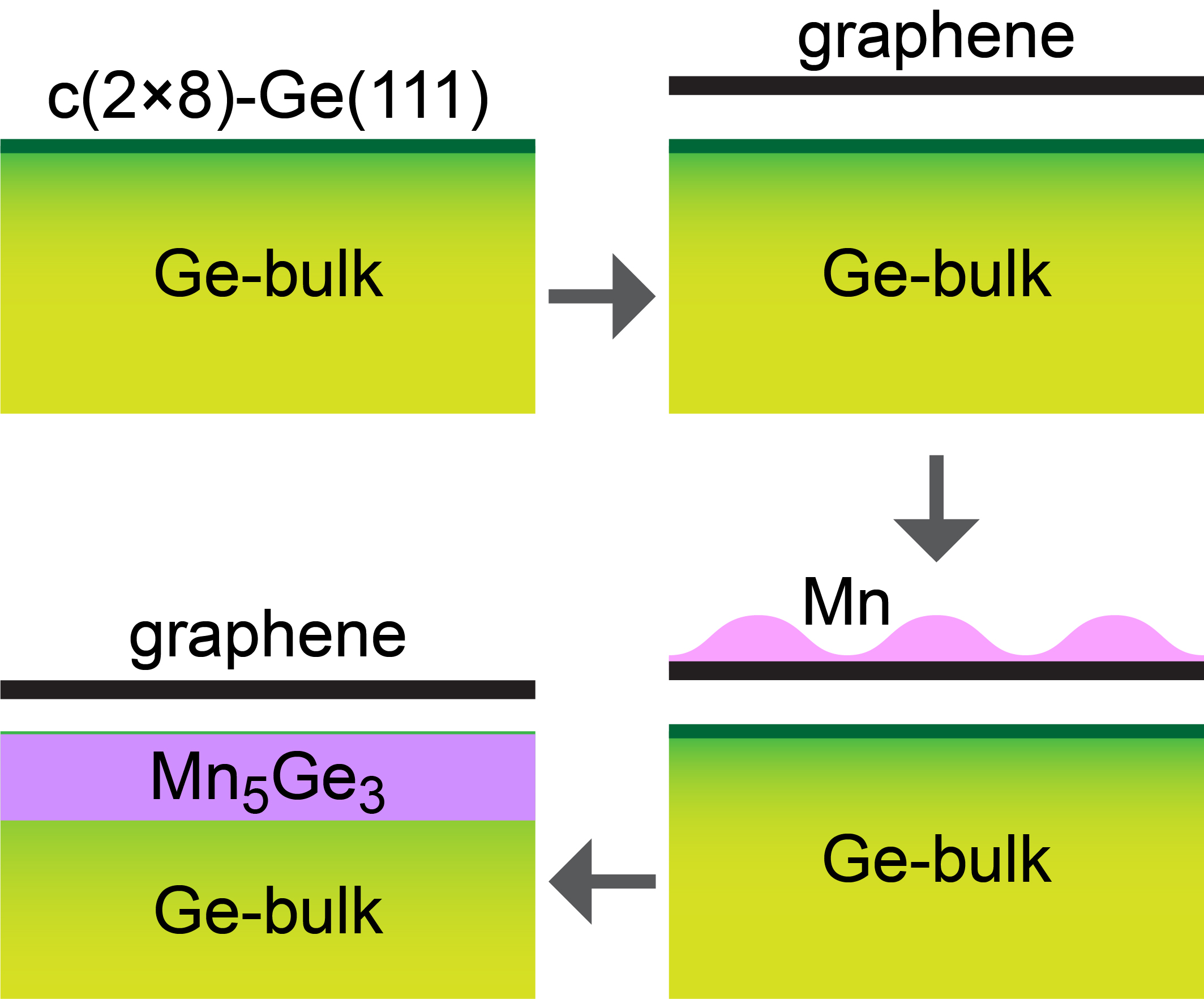}
\caption{Two-steps preparation procedure for the gr-Mn$_5$Ge$_3$(0001) system: (i) graphene on Ge(111) is grown via CVD or MBE; (ii) atomically sharp gr-Mn$_5$Ge$_3$(0001) interface is formed via intercalation of the pre-deposited thin layer of Mn on gr-Ge(111).}
\label{scheme}
\end{figure}

The preparation of the gr-Mn$_5$Ge$_3$(0001) interface can be accomplished in two steps (Figure~\ref{scheme}). On the first step graphene is grown via chemical vapour deposition (CVD) or molecular beam epitaxy (MBE) on semiconducting Ge(111). As was demonstrated earlier~\cite{Lee:2014dv}, in this case graphene is aligned on this surface with the graphene zig-zag edges along the Ge$\langle1\bar{1}0\rangle$ and the arm-chair edges along Ge$\langle11\bar{2}\rangle$. On the second preparation step a thin layer of FM metal Mn$_5$Ge$_3$(0001) can be grown under graphene via well established intercalation procedure, where thin Mn layer is deposited on graphene and then subsequently annealed at the elevated temperature. It is well known that FM Mn$_5$Ge$_3$(0001) can be epitaxially grown on Ge(111) in the temperature range $300-650^\circ$\,C~\cite{Zeng:2003,Dedkov:2009a,Grytzelius:2011dt} and it has relatively high Curie temperature of $296$\,K, which can be increased well above room temperature via chemical doping~\cite{Gajdzik:2000ea,Slipukhina:2009,Stojilovic:2013bx,Petit:2015dl}. Also previous experiments demonstrate the gr/Ge interface is stable at temperatures, at least, up to $800^\circ$\,C~\cite{Kiraly:2015kaa,Tesch:2017gm,Tesch:2018hm}. Thus, the proposed preparation routine can lead to the epitaxial heterostructure consisting of 2D graphene, FM Mn$_5$Ge$_3$, and semiconducting Ge. 

\begin{figure}[t]
\centering
\includegraphics[width=0.57\columnwidth]{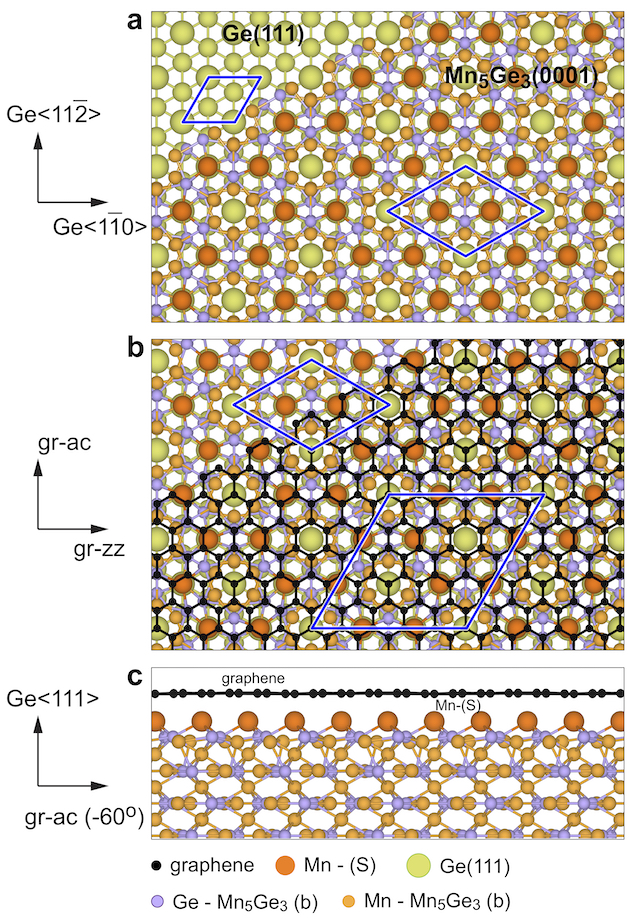}
\caption{Top views of the crystallographic structure of (a) Mn$_5$Ge$_3$(0001)/Ge(111) and (b) gr/Mn$_5$Ge$_3$(0001)/Ge(111) (structure B). (c) Side view of the fully relaxed optimized gr-Mn$_5$Ge$_3$(0001) interface (structure B). Blue rhombuses mark unit cells of $(1\times1)$-Ge(111), $(\sqrt{3}\times\sqrt{3})R30^\circ$-Mn$_5$Ge$_3$(0001), and $(5\times5)$-gr/Mn$_5$Ge$_3$(0001), respectively. Crystallographic directions of Ge-bulk as well as graphene arm-chair (ac) and zig-zag (zz) edges are marked on the left hand side of the image. Side view in (c) is taken for the view direction perpendicular to the main diagonal of the big rhombus in (b). Corresponding colored circles are used for carbon atoms in graphene and for Mn and Ge atoms in bulk (b) and at the interface (S).}
\label{structure}
\end{figure}

The crystallographic structure, top and side views, of the respective interfaces discussed above are shown in Figure~\ref{structure}. FM Mn$_5$Ge$_3$ has a hexagonal structure and it epitaxially grows on the Ge(111) surface forming the $(\sqrt{3}\times\sqrt{3})R30^\circ$ structure with respect to $(1\times1)$-Ge(111) (Figure~\ref{structure}a). As described earlier, the gr-Mn$_5$Ge$_3$(0001) interface is formed via intercalation of a Mn layer in the gr-Ge(111) system. Two possible arrangements of C-atoms on Mn$_5$Ge$_3$(0001) are considered, structure B and structure T (Figure\,S1).  Both structures are very similar with respect to their energetical stability (the difference in the interaction energy of graphene with substrate is of only $7$\,meV/C-atom). It is found that the discussed effects are valid for both structures and they are robust to the relative orientation of the layers at the gr-Mn$_5$Ge$_3$ interface.  The resulting most energetically favourable interface (structure B) has the structure presented in Figure~\ref{structure}b, where the corresponding orientation of graphene with respect to the Ge(111) surface is preserved in order to avoid any layers rotations. For the formed gr-Mn$_5$Ge$_3$(0001) interface the unit cell of the obtained structure has a periodicity of 5 unit cells of graphene and this unit cell (as well as the primitive unit cell of graphene) is rotated by $30^\circ$ with respect to the unit cell of the underlying Mn$_5$Ge$_3$. The side view of the fully relaxed gr-Mn$_5$Ge$_3$(0001) interface is shown in Figure~\ref{structure}c with the mean distance between graphene and the top Mn layer of $2.144$\,\AA. Considering the fact that the surface of Mn$_5$Ge$_3$ is Mn-terminated with the relatively small density of these atoms in the surface layer one can expect the appearance of the substantial corrugation of a graphene layer in this system. However, our calculations show that here graphene is quite flat with the corrugation of only $0.122$\,\AA\ and the interaction energy of graphene with the Mn$_5$Ge$_3$(0001) surface is $149.5$\,meV/C-atom, which places this system to the sub-class of a \textit{strongly} interacting graphene with metallic substrate. The relatively small corrugation of graphene in the considered lattice mismatched system can be connected with the variation of the local environment for carbon atoms on the distance of not more than 2 carbon rings. The strong $sp^2$ character of carbon atoms in graphene and the rigidity of the $\sigma$ bonds prevents the strong long-distance corrugation effect for the graphene layer. The flatness of graphene in this system might be crucial for the use of this interface as a spin filter thus avoiding the strong charge and spin scattering due to the interface corrugation on the atomic level. The simulated scanning tunnelling microscopy (STM) images of the Mn$_5$Ge$_3$(0001) surface before and after graphene adsorption are shown in Figure\,S2 and can be used as a reference for the future experimental studies. The theoretical STM results for the clean surface are in very good agreement with the available experimental and theoretical data~\cite{Zeng:2003,Grytzelius:2011dt,Zhang:2013cq} confirming the correctness of the approach used in the present study.

\begin{figure}[t]
\centering
\includegraphics[width=0.57\columnwidth]{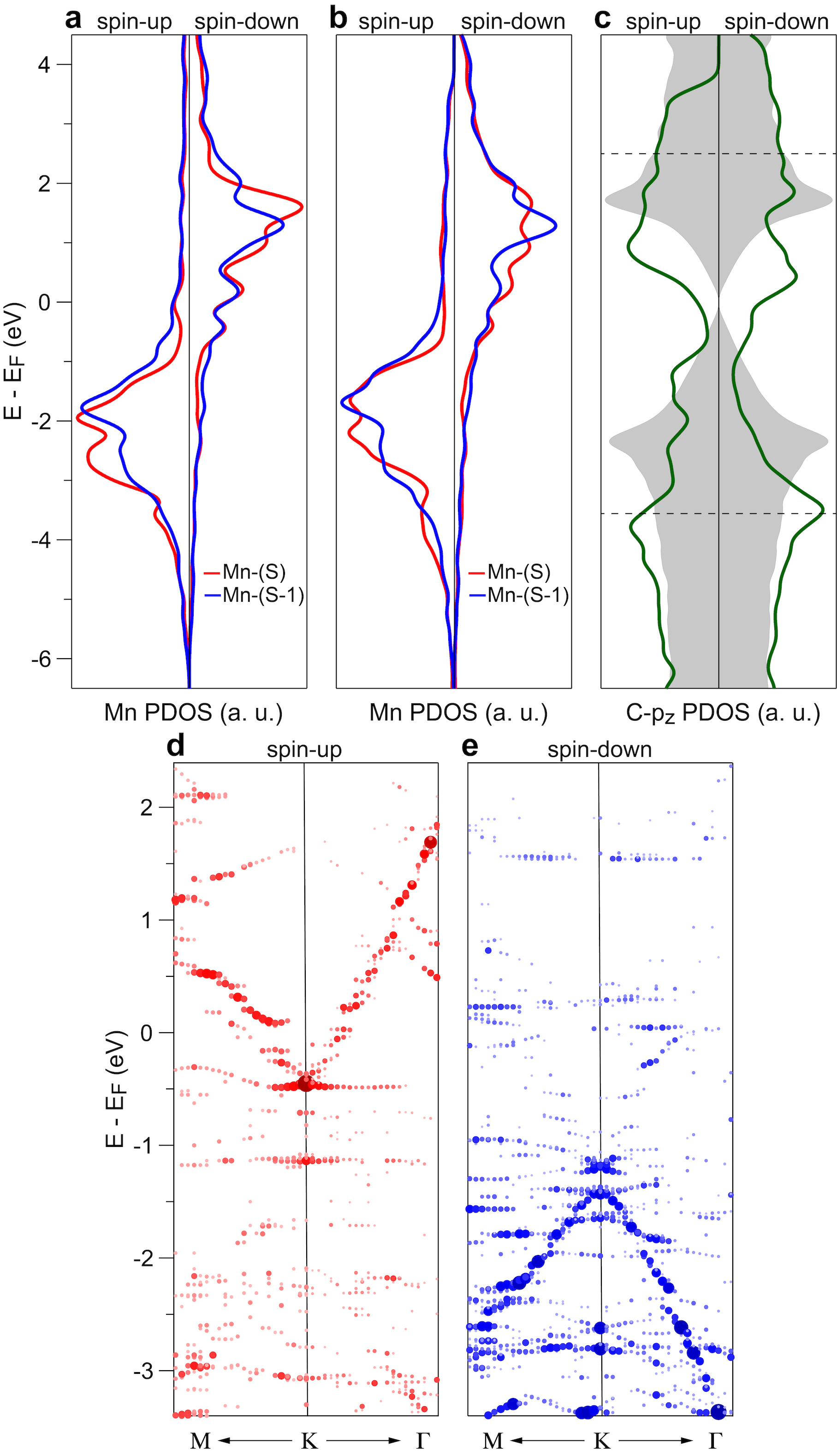}
\caption{Spin-resolved Mn-atom projected PDOS for (a) the clean Mn$_5$Ge$_3$(0001) surface and (b) the gr-Mn$_5$Ge$_3$(0001) interface (structure B). (c) Spin-resolved C-atom projected $p_z$ PDOS for $(5\times5)$-gr/Mn$_5$Ge$_3$(0001) (structure B). Grey shadowed area in (c) shows the C\,$p_z$ PDOS for the free-standing graphene in the same $(5\times5)$ unit cell. (d,e) Spin-resolved band structure obtained after unfolding procedure and presented around the $\mathrm{K}$ point of the $(1\times1)$ graphene-derived Brillouin zone for the $(5\times5)$-gr/Mn$_5$Ge$_3$(0001) system (structure B) for spin-up and spin-down channels, respectively. The size of the point gives the information about the number of primitive cell bands crossing particular ($k$,$E$) in the unfolded procedure, i. e. the partial density of states at ($k$,$E$).  Energy range in (d,e) corresponds to the one marked by the dashed lines in (c).}
\label{DOS_bands}
\end{figure}

Before adsorption of graphene on Mn$_5$Ge$_3$(0001) the density of states of this material is strongly exchange split (Figure~\ref{DOS_bands}a and Figure\,S3). The magnetic moment of the surface Mn-(S) atoms is $3.48\mu_B$, which is larger compared to the value of $3.11\mu_B$ for the subsurface Mn-(S-1) atoms, that can be explained by the reduced number of neighbours at the surface and the respective narrowing of the Mn $3d$ band. Adsorption of graphene leads to the substantial reduction of the magnetic moment of Mn-(S) and Mn-(S-1) to $3.11\mu_B$ and $2.98\mu_B$, respectively, which is reflected as the decrease of the exchange splitting of the respective states in Mn-atom projected partial DOS (PDOS) (Figure~\ref{DOS_bands}b). At the same time the $\pi$ states of graphene become spin polarized due to the presence of the FM Mn$_5$Ge$_3$ substrate  (Figure~\ref{DOS_bands}c). The calculated DOS for structure T and magnetic moments for all atoms in the studied systems can be found in Figure\,S4, in Table\,S5, and in the Supplementary data files.

The space-, energy, and $k$-vector overlap of the valence band states of graphene and Mn$_5$Ge$_3$ leads to the effective hybridization of these states. At the same time the spin character of these states is also conserved in this process. This can be clearly seen in Figure\,S6 where large-energy scale band structures of the gr/Mn$_5$Ge$_3$(0001) interface obtained after unfolding procedure are shown (for comparison reasons, in case of structure B, DFT calculations were performed with and without inclusion of the spin-orbit interaction). Considering the DOS of Mn$_5$Ge$_3$ (Figure~\ref{DOS_bands}a,b and shadowed areas in Figure\,S6) we can clearly see that spin-polarized graphene $\pi$ bands hybridize with the Mn-atoms projected valence band states of the only respective spin character. This effect leads to the ``huge'' band gaps for both graphene $\pi$ bands of different spin characters around the $K$ point of the graphene-derived $(1\times1)$ Brillouin zone: for spin-up states this gap is between $E-E_F\approx-3.8\mathrm{eV}\ldots-0.5\mathrm{eV}$ and for spin-down states this gap is between $E-E_F\approx-1.4\mathrm{eV}\ldots+2.1\mathrm{eV}$. At the same time the effect of hybridization leads to the appearance of the so-called interface states of the mixed graphene-$\pi$ and Mn\,$3d$ character, which manifest themselves as a series of flat bands of the respective spin character in the energy range $E-E_F\approx-3.8\mathrm{eV}\ldots+2.1\mathrm{eV}$. The similar effect of the interface states formation was observed earlier for other gr-FM interfaces, gr/Ni(111) and gr/Co(0001)~\cite{Bertoni:2004,Karpan:2008,Dedkov:2010jh,Voloshina:2011NJP,Voloshina:2011wa}.

The effect of the spin-conservation during hybridization of the valence band states at the gr-Mn$_5$Ge$_3$(0001) interface leads to the interesting results, which have dramatic implications on the spin-filtering properties of this interface. Analysis of the zoomed region around the $K$ point and in the vicinity of $E_F$ of the band structure of this system showed in Figure~\ref{DOS_bands}d,e demonstrates that only spin-up electrons in graphene retain the Drirac-electron-like character with the linear energy dispersion for these electronic states. There is a clear anisotropy for the band dispersion along the $\Gamma-\mathrm{K}$ and $\mathrm{M}-\mathrm{K}$ directions and the linear fit of the graphene $\pi$ band around $E_F$ gives the Fermi velocity of $v_F=0.(92\pm0.01)\times10^6\mathrm{m/s}$ and $v_F=0.(71\pm0.01)\times10^6\mathrm{m/s}$ for $\Gamma-\mathrm{K}$ and $\mathrm{M}-\mathrm{K}$, respectively, which is very close to the value characteristic for free-standing graphene and which is much higher compared to the value which was extracted for the strongly hybridized states formed at the gr/Co(0001) interface~\cite{Usachov:2015kr}. Moreover, this spin-up state has a linear dispersion in the energy range of $\approx1\mathrm{eV}$ around $E_F$ that significantly simplifies the energy position control for this state. For spin-down electrons, there is a relatively large gap between $E-E_F\approx-1.2\mathrm{eV}\ldots+0.1\mathrm{eV}$ at the $K$ point. Above and below this energy window a series of the interface states is formed which was described above. The carriers on these states have very small group velocity due to the relatively large Mn\,$3d$ character. All these factors allow us to conclude that the contact of graphene with FM-M$_5$Ge$_3$ will have a strong influence on its spin-transport properties.

In conclusion, we have discovered a new interface of graphene with strong ferromagnetic material Mn$_5$Ge$_3$, which valence band states are strongly exchange split. This system can be easily fabricated using two steps preparation routine: CVD or MBE of graphene on the catalytically active Ge(111) surface and then intercalation of Mn in the gr-Ge(111) interface with the formation of the ordered gr-Mn$_5$Ge$_3$(0001) interface. Our systematic large-scale DFT calculations for this interface demonstrate that graphene is relatively strongly bonded to the FM Mn$_5$Ge$_3$ substrate that leads to the substantial exchange splitting of the graphene $\pi$ states. It is found that the effect of the spin-conservation during hybridization of the graphene $\pi$ and Mn\,$3d$ valence band states leads to the appearance of the large band gaps in the energy dispersion of the graphene $\pi$ states for both spin channels. The unique combination of all factors leads to the observation of the only spin-up electrons of graphene at the $\mathrm{K}$ point and at $E_F$ which conserve the Dirac-electron-like character. At the same time for the spin-down electrons the band gap is found at $E_F$. We expect that our study will stimulate future experimental effort to create graphene-semiconductor and graphene-FM structures and we can expect that the possible 2D\,graphene -- FM\,Mn$_5$Ge$_3$ -- semiconducting\,Ge systems will be used in the charge and spin transport studies and applications.

The North-German Supercomputing Alliance (HLRN) is acknowledged for providing computer time.

\providecommand{\latin}[1]{#1}
\makeatletter
\providecommand{\doi}
  {\begingroup\let\do\@makeother\dospecials
  \catcode`\{=1 \catcode`\}=2 \doi@aux}
\providecommand{\doi@aux}[1]{\endgroup\texttt{#1}}
\makeatother
\providecommand*\mcitethebibliography{\thebibliography}
\csname @ifundefined\endcsname{endmcitethebibliography}
  {\let\endmcitethebibliography\endthebibliography}{}

\clearpage

\clearpage

\clearpage

\clearpage
\noindent
\textbf{Supporting Information for: ``Dirac-Electron Behavior for the Spin-Up Electrons in the Strongly Interacting Graphene on Ferromagnetic Mn$_5$Ge$_3$''}

\section*{Computational Details}

Spin-polarised DFT calculations based on plane-wave basis sets of $500$\,eV cutoff energy were performed with the Vienna \textit{ab initio} simulation package (VASP)~\cite{Kresse:1994,Kresse:1996}. The Perdew-Burke-Ernzerhof (PBE) exchange-correlation functional~\cite{Perdew:1996} was employed. The electron-ion interaction was described within the projector augmented wave (PAW) method~\cite{Blochl:1994} with C ($2s$,$2p$), Ge ($4s$,$4p$), and Mn ($3d$,$4s$) states treated as valence states. The Brillouin-zone integration was performed on $\Gamma$-centred symmetry reduced Monkhorst-Pack meshes using a Methfessel-Paxton smearing method of first order with $\sigma = 0.15$\,eV, except for the calculations of total energies. For these calculations, the tetrahedron method with Bl\"ochl corrections~\cite{Blochl:1994vg} was employed. The $k$ mesh for sampling the supercell Brillouin zone are chosen to be as dense as $18\times18\times24$,  $18\times18\times1$ and $12\times12\times1$ in the case of Mn$_5$Ge$_3$ bulk, Mn$_5$Ge$_3$ (0001) surface and gr/Mn$_5$Ge$_3$(0001) interface, respectively. Dispersion interactions were considered by adding a $1/r^6$ atom-atom term as parameterised by Grimme (``D2'' parameterisation)~\cite{Grimme:2006}. For comparisons reasons, some calculations (results are marked in the respective figures) were performed with spin-orbit interaction taken into account. The calculated band structures were unfolded (where necessary) to the graphene ($1\times1$) primitive unit cells according to the procedure described in Refs.~\citenum{Medeiros:2014ka}, \citenum{Medeiros:2015ks}  with the code BandUP.

Bulk Mn$_5$Ge$_3$ is a ferromagnetic intermetallic compound. It crystallises in the hexagonal $D8_8$ structure (prototype Mn$_5$Si$_3$ and space group $P6_3/mcm$) and has lattice parameters of $a=7.184$\,\AA\ and $c=5.053$\,\AA~\cite{Forsyth:1990ia}. The hexagonal cell contains $10$ Mn (two sub-lattices, Mn1 and Mn2) and $6$ Ge atoms. Four atomic planes are situated perpendicular to the axis $z$. Planes $z=0$ and $z=1/2$ contain only Mn1-atoms, planes $z=1/4$ and $z=3/4$ contain Ge- and Mn2-atoms. Mn1-atoms are surrounded by $6$ Ge-atoms and Mn2-atoms are surrounded by $5$ Ge-atoms. The atomic positions are: Mn1 in 4(d) site: $\pm(1/3,2/3,0;\,2/3,1/3,1/2)$; Mn2 in 6(g) site: $\pm(x_1,0,1/4;\,0,x_1,1/4;\,-x_1,-x_1,1/4)$ with $x_1=0.2397$, Ge in 6(g) site: $\pm(x_2,0,1/4;\,0,x_2,1/4;\,-x_2,-x_2,1/4)$ with $x_2=0.6030$. The optimised lattice parameters ($a=7.145$\,\AA, $c=4.976$\,\AA, $x_1=0.2435$, $x_2=0.6059$) were used in order to construct models for the (0001) surface as well as interface with graphene.

The Mn$_5$Ge$_3$(0001) surface was modelled by a slab containing $12$ atomic layers and a vacuum gap of approximately $23$\,\AA. The atoms of the $8$ bottom layers were fixed at their bulk positions during the structural optimisation procedure, whereas the positions ($x$, $y$, $z$-coordinates) of all other atoms were fully relaxed until forces became smaller than $0.02$\,eV\,\AA$^{-1}$.

To model the gr/Mn$_5$Ge$_3$(0001) interface, a $\left(\sqrt{3}\times\sqrt{3}\right)R30^\circ$ structure was formed with respect to Mn$_5$Ge$_3$(0001) slab and its top side was covered by a graphene layer consisting of $(5\times5)$ unit cells. Thus, the resulting supercell has the size $(12.375\,\textrm{\AA}\times12.375\,\textrm{\AA})$ and consists of $50$ C-atoms, $90$ Mn-atoms and $54$ Ge-atoms. In the resulting structure the graphene lattice was stretched by $0.6$\% compared to the equilibrium lattice constant of $2.461$\,\AA. Two possible arrangements of C-atoms above the Mn$_5$Ge$_3$(0001) surface were considered - structure B and  structure T. In structure B (which is energetically more favourable one) all Mn atoms of the interface layer are bridged by the C-atoms of the graphene layer. Structure T is obtained by shift of a graphene layer in the $(x,y)$ plane in such a way, that C atoms of graphene occupy all possible high-symmetry adsorption positions with respect to the interface Mn-layer. As for the clean surface, the atoms of the $8$ bottom layers were fixed at their bulk positions during the structural optimization procedure, whereas the positions ($x$, $y$, $z$-coordinates) of all other atoms were fully relaxed until forces became smaller than $0.02$\,eV\,\AA$^{-1}$.

\fontsize{12}{9}\selectfont

\normalsize

\clearpage

\noindent
\textbf{List of Tables and Figures}
\newline
\newline
\noindent
Fig.\,S1: Top and side views of (a) structure B and (b) structure T. Corresponding colored circles are used for carbon atoms in graphene and for Mn and Ge atoms in bulk (b) and at the interface (S).
\newline
\noindent
Fig.\,S2: Calculated STM images for (a) the clean Mn$_5$Ge$_3$(0001) surface and for the gr/ Mn$_5$Ge$_3$(0001) system -- (b) structure B and (c) structure T. Bias voltages are marked in every image.
\newline
\noindent
Fig.\,S3: Spin resolved band structures of (a) bulk Mn$_5$Ge$_3$ and (b) Mn$_5$Ge$_3$(0001) surface. The respective Brillouin zones are shown on the right-hand side for every panel.
\newline
\noindent
Fig.\,S4: Spin-resolved Mn-atom projected PDOS for (a) the clean Mn$_5$Ge$_3$(0001) surface and (b) the gr/Mn$_5$Ge$_3$(0001) interface (structure T). (c) C-atom projected $p_z$ PDOS for $(5\times5)$-gr/Mn$_5$Ge$_3$(0001) (structure T). Grey shadowed area in (c) show the C\,$p_z$ PDOS for the free-standing graphene in the same $(5\times5)$ unit cell.
\newline
\noindent
Tab.\,S5: Parameters, interaction energy, distances between layers, and magnetic moments, for the Mn$_5$Ge$_3$(0001) surface and gr/Mn$_5$Ge$_3$(0001) (structures B and T). All distances are marked according to Fig.\,S1. The respective structures can be visualized using software VESTA (http://jp-minerals.org/vesta/en/); instructions are placed as a header in every txt-file. Structures files are available upon request.
\newline
\noindent
Fig.\,S6: Large-energy scale spin-resolved band structures for the gr/Mn$_5$Ge$_3$(0001) system shown for structure B without (a) and with (b) inclusion of the spin-orbit interaction. In (c) the large-energy scale spin-resolved band structure (without inclusion of the spin-orbit interaction) for structure T is shown. Data in (b) and (c) are shown only for $\Gamma-\mathrm{K}$ direction. In (a), the discussed in the main text band gaps for the graphene $\pi$ states are marked with $E_g$ for spin-up and spin-down channels..

\clearpage
\begin{figure}[t]
\centering
\includegraphics[width=0.9\textwidth]{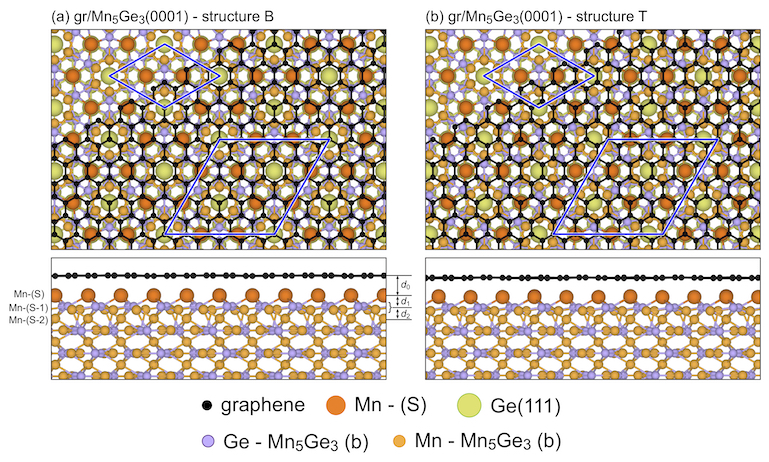}
\label{Fig_S1_strBstrT}
\end{figure}
\noindent
Fig.\,S1: Top and side views of (a) structure B and (b) structure T. Corresponding colored circles are used for carbon atoms in graphene and for Mn and Ge atoms in bulk (b) and at the interface (S).

\clearpage
\begin{figure}[t]
\centering
\includegraphics[width=0.66\textwidth]{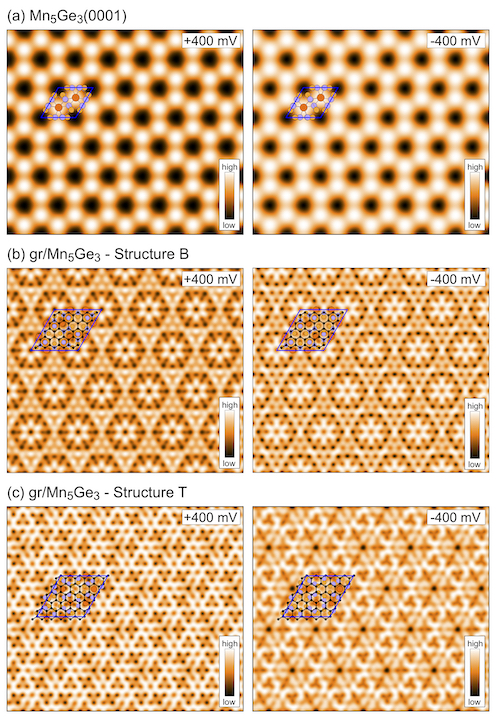}\\
\label{Fig_S2_STM}
\end{figure}
\noindent
Fig.\,S2: Calculated STM images for (a) the clean Mn$_5$Ge$_3$(0001) surface and for the gr/Mn$_5$Ge$_3$(0001) system -- (b) structure B and (c) structure T. Bias voltages are marked in every image.

\clearpage
\begin{figure}[t]
\centering
\includegraphics[width=0.7\textwidth]{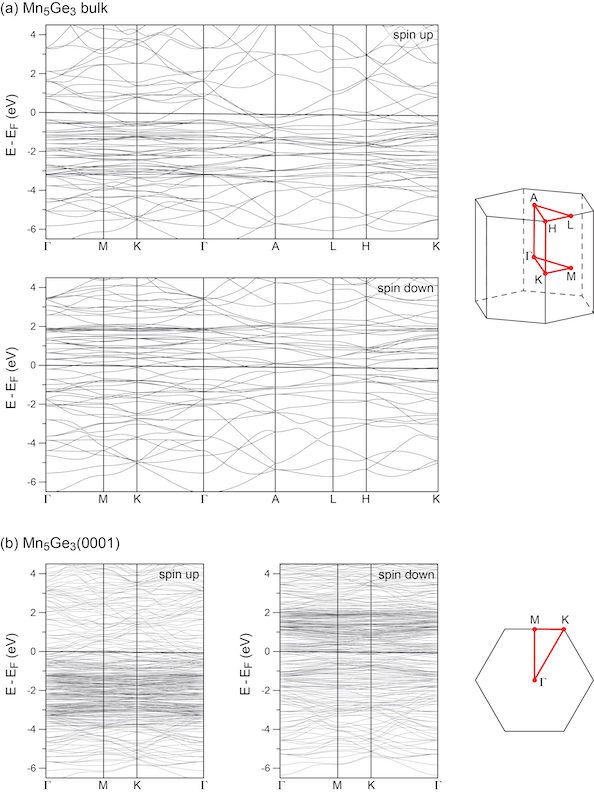}\\
\label{Fig_S3_bands_Mn5Ge3}
\end{figure}
\noindent
Fig.\,S3: Spin resolved band structures of (a) bulk Mn$_5$Ge$_3$ and (b) Mn$_5$Ge$_3$(0001) surface. The respective Brillouin zones are shown on the right-hand side for every panel.

\clearpage
\begin{figure}[t]
\centering
\includegraphics[width=0.9\textwidth]{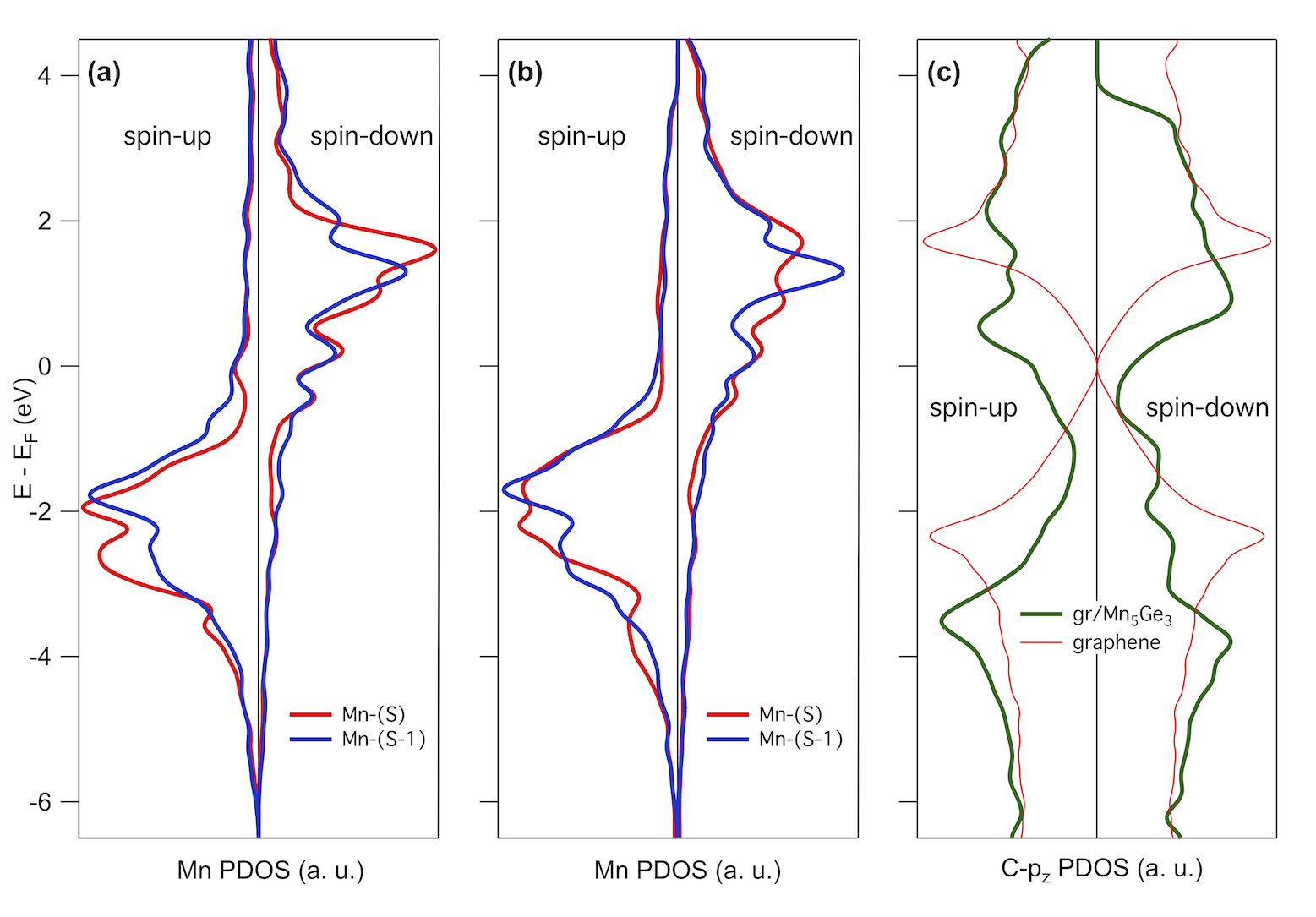}\\
\label{Fig_S4_DOS_T}
\end{figure}
\noindent
Fig.\,S4: Spin-resolved Mn-atom projected PDOS for (a) the clean Mn$_5$Ge$_3$(0001) surface and (b) the gr/Mn$_5$Ge$_3$(0001) interface (structure T). (c) C-atom projected $p_z$ PDOS for $(5\times5)$-gr/Mn$_5$Ge$_3$(0001) (structure T). Thin line in (c) showsthe C\,$p_z$ PDOS for the free-standing graphene in the same $(5\times5)$ unit cell.

\clearpage

\noindent Table\,S5: Parameters, interaction energy, distances between layers, and magnetic moments, for the Mn$_5$Ge$_3$(0001) surface and gr/Mn$_5$Ge$_3$(0001) (structures B and T). All distances are marked according to Fig.\,S1. The respective structures can be visualized using software VESTA (http://jp-minerals.org/vesta/en/); instructions are placed as a header in every txt-file. Structures files are available upon request.
\bigskip

\begin{tabular}{|l |c|c|c|}
\hline
  &Mn$_5$Ge$_3$(0001) &\multicolumn{2}{c|}{gr/Mn$_5$Ge$_3$}\\
 \cline{3-4}
  &                   &Structure B & Structure T \\ 
\hline
E$_\textrm{int}$/meV         &         &$-149.5$   &$-142.6$  \\
$d_0$/\AA      &         &$2.144$    &$2.137$ \\
$d_1$/\AA   &$1.286$  &$1.377$    &$1.378$ \\
$d_2$/\AA &$1.232$  &$1.144$    &$1.145$ \\
$m_\textrm{Mn(S)}/\mu_B$     &see Structure-surf.txt  &see Structure-B.txt   &see Structure-T.txt  \\
$m_\textrm{Mn(S-1)}/\mu_B$   &see Structure-surf.txt   &see Structure-B.txt  &see Structure-T.txt \\
$m_\textrm{C}/\mu_B$         &see Structure-surf.txt   &see Structure-B.txt  &see Structure-T.txt \\
\hline
\end{tabular}\\
\newline
\noindent

\clearpage
\begin{figure}[t]
\centering
\includegraphics[width=0.66\textwidth]{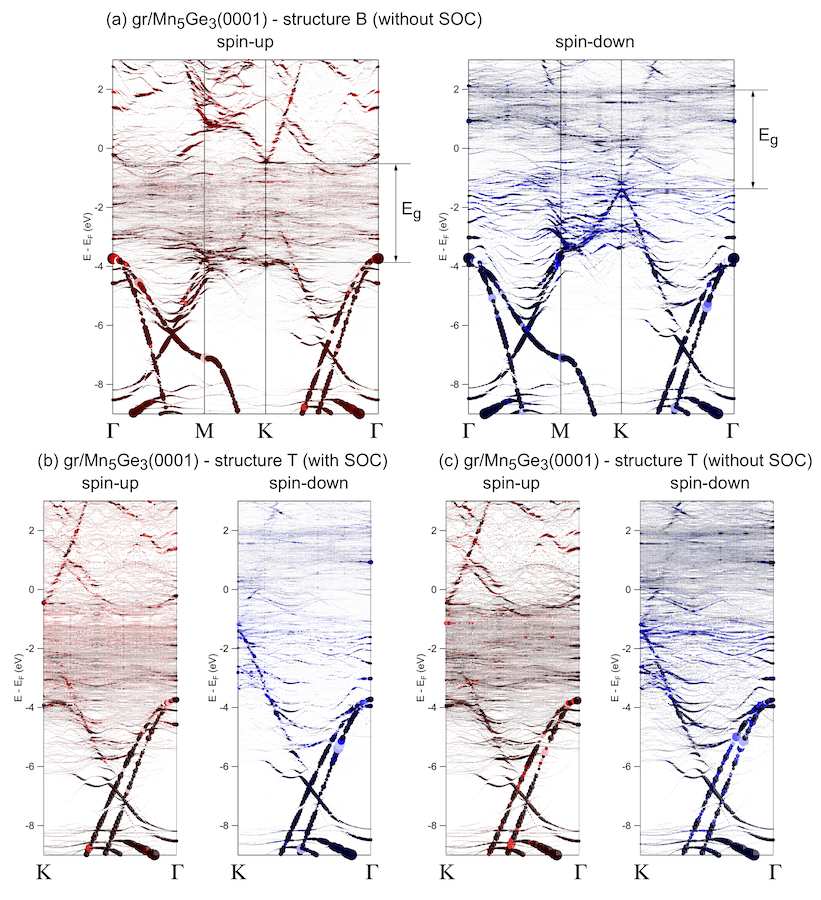}\\
\label{Fig_S6_bands_strB_strT}
\end{figure}
\noindent
Fig.\,S6: Large-energy scale spin-resolved band structures for the gr/Mn$_5$Ge$_3$(0001) system shown for structure B without (a) and with (b) inclusion of the spin-orbit interaction. In (c) the large-energy scale spin-resolved band structure (without inclusion of the spin-orbit interaction) for structure T is shown. Data in (b) and (c) are shown only for $\Gamma-\mathrm{K}$ direction. In (a), the discussed in the main text band gaps for the graphene $\pi$ states are marked with $E_g$ for spin-up and spin-down channels.

\end{document}